# Neural Network-Assisted End-to-End Design for Dispersive Full-Parameter Control of Meta-Optics

Hanbin Chi[1,#], Yueqiang Hu[1,2,3,#,*], Xiangnian Ou[1], Yuting Jiang[1], Dian Yu[1], Shaozhen Lou[1], Quan Wang[1], Qiong Xie[1], Cheng-Wei Qiu[4,*], and Huigao Duan[1,2,*]

[1]National Research Center for High-Efficiency Grinding, College of Mechanical and Vehicle Engineering, Hunan University, Changsha 410082, P. R. China.

[2]Greater Bay Area Institute for Innovation, Hunan University, Guangzhou 511300, China.

[3]Advanced Manufacturing Laboratory of Micro-Nano Optical Devices, Shenzhen Research Institute, Hunan University, Shenzhen 518000, China.

[4]Department of Electrical and Computer Engineering, National University of Singapore, Singapore 117583, Singapore.

[#]These authors contributed equally to this work.

[*]Corresponding authors. E-mail: huyq@hnu.edu.cn; chengwei.qiu@nus.edu.sg; duanhg@hnu.edu.cn

# Abstract

Flexible control light field across multiple parameters is the cornerstone of versatile and miniaturized optical devices. Metasurfaces, comprising subwavelength scatterers, offer a potent platform for executing such precise manipulations. However, the inherent mutual constraints between parameters of metasurfaces make it challenging for traditional approaches to achieve full-parameter control across multiple wavelengths. Here, we propose a universal end-to-end inverse design framework to directly optimize the geometric parameter layout of meta-optics based on the target functionality of full-parameter control across multiple wavelengths. This framework employs a differentiable forward simulator integrating a neural network-based dispersive full-parameter Jones matrix and Fourier propagation to facilitate gradient-based optimization. Its superiority over sequential forward designs in dual-polarization channel color holography with higher quality and tri-polarization three-dimensional color holography with higher multiplexed capacity is showcased. To highlight the universality, we further present polarized spectral multi-information processing with six arbitrary polarizations and three wavelengths. This versatile, differentiable, system-level design framework is poised to expedite the advancement of meta-optics in integrated multi-information display, imaging, and communication, extending to multi-modal sensing applications.

**Introduction**

The spatial light field is fully characterized by amplitude, phase, and polarization of different wavelength components at each spatial location[1,2]. Manipulating these individual or multiple parameters constitutes the foundation of all optical elements and systems[3,4]. However, achieving independent control over multiple parameters or even full-parameter control is challenging with traditional refractive optical elements. Metasurfaces, constructed with subwavelength scatters, offer a potent platform to arbitrarily modulate parameters of light in ultra-thin planar elements due to their vast design flexibility[5,6]. Fig. 1a categorizes multi-parameter light field control with single-layer metasurfaces using a three-axis coordinate system representing wavelength, complex amplitude, and polarization. By controlling a single parameter, metasurfaces have been developed for functions such as beam steering[7,8], metalenses[9,10], meta-holograms[11,12] and polarization converters[13,14]. Concurrently manipulating multiple parameters of light enables integrated multifunctionality within an ultra-compact form factor, effectively leveraging the benefits of metasurfaces[15-17].

Extensive research has focused on multi-parameter control at a single wavelength, utilizing dual-parameter control through pairwise combinations of amplitude, phase, and polarization to achieve complex amplitude elements[18-20], polarization-assisted nano-printing[21-23] and dual-channel multiplexing[24-27]. Further advancements in polarization engineering, based on Jones matrix, have facilitated precise manipulation of amplitude or phase across three polarization components[28-30] and complex-amplitude over two[31-34] or three[35,36] polarization components, indicating that full-parameter manipulation at a single wavelength has been thoroughly explored. In the domain of multi-wavelength design, controlling a single parameter across multiple wavelengths has advanced the development of achromatic metalenses[35,36] and color nano-printing[39]. For multi-parameter control across multiple wavelengths, interleaved meta-atoms or K-space engineering have achieved full-color complex amplitude[40,41], dual-polaraization[42], and full-polarization holography[43,44], although limited by spatial resolution and secondary diffraction effects. As functionality becomes more complex and performance demands increase, identifying suitable meta-atoms for dispersion-engineered multi-parameter control becomes difficult due to the complex crosstalk between parameters at different wavelength channels. Deriving a universal design solution for full-parameter control of spatial light field through forward semi-analytical methods is exceedingly challenging. Inverse design, employing deep learning and optimization algorithms, has presented significant advantages for polarization and wavelength multiplexing with multiple[45,46] and freeform[47] meta-atoms, while facing substantial challenges in fabrication and model training. Moreover, these attempts typically only consider partial parameters of light, neglecting the potential for full parameter control with single layer meta-optics across multiple wavelengths, as outlined at the highest level in Fig. 1a. Consequently, developing an inverse design framework for full-parameter control of meta-optics with a simplified configuration is crucial.

In this work, we propose a comprehensive end-to-end inverse design framework for full-parameter light field manipulation using non-interleaved dielectric meta-optics，representing the highest level in Fig. 1a. This framework integrates a neural network-based dispersive full-parameter Jones matrix and Fourier propagation to construct a differentiable forward simulator, facilitating gradient-based optimization. It enables direct determination of geometric parameters for meta-optics fabrication based on multi-objective functions that considers modulations in amplitude, phase, and polarization across multiple wavelengths at different spatial locations. Our framework achieves superior experimental performance in dual-polarization color holography compared with sequential forward design. To demonstrate its universality, we showcase several cases using non-interleaved meta-optics that are impractical by sequential forward design. For the first time experimentally, we realize three polarization channels for three-dimensional color holography including independent nine channels. Furthermore, we demonstrate arbitrary polarized spectral multi-information processing under six arbitrary polarizations and three wavelengths for meta-hologram and metalens. This multifunctional, differentiable, system-level end-to-end inverse design framework enables flexible customization of full-parameter modulation of spatial light field with meta-optics, paving the way for the development of meta-optics in integrated multi-information display, imaging, communication, and even multimodal sensing applications.

## Results

**Neural network-assisted end-to-end inverse design framework**

As shown in the upper of Fig. 1b, end-to-end inverse design requires identifying the optimal geometric parameters layout of meta-optics for full-parameter control of spatial light wave. This task can be formulated as a multi-objective optimization problem:

$$\arg\min_{S} Loss = \arg\min_{S} \sum_{\lambda,p,z} | E_{target}(\lambda, p, z) - E_{ms}(S; \lambda, p, z) | \qquad (1)$$

where $E_{target}$ represents the desired spatial light field, characterized by its amplitude ($A$), phase ($\varphi$), polarization ($p$), wavelength ($\lambda$) at each spatial location ($x, y, z$). $E_{ms}$ denotes the modulated field by meta-optics with limited geometric parameters $S \in [S_{min}, S_{max}]$. The difference between $E_{target}$ and $E_{ms}$ is defined as the loss function, serves as the objective function in end-to-end inverse design. This high-dimensional optimization problem, constrained by multiple objectives, is addressed by iteratively minimizing the loss function though gradient-based optimization. Below Fig. 1b, the modules of end-to-end inverse design framework incorporates a forward simulator with a dispersive full-parameter Jones matrix and Fourier propagation, an objective function, gradient-based optimization, and an iterative loop.

To obtain the gradient information of $E_{target}$ with respect to $S$, we have modeled a differentiable forward simulator for meta-optics design. The component unit of the meta-optics, referred to as meta-atom, is depicted in an inset of Fig. 1b. Each meta-atom consists of a single titanium dioxide ($TiO_2$) rectangular unit on a silicon dioxide ($SiO_2$) substrate with a subwavelength square period constant P = 400 nm, a fixed height h = 1000 nm, and three variable geometric parameters: length ($L$), width ($W$), and rotation angle ($\theta$). Meta-optics is defined by the set of geometric parameters $S$ of the meta-atom at $xy$ coordinate, represented as $S = \{L, W, \theta | (x, y)\}$. The dispersion and polarization characteristics of the birefringent meta-optics can be expressed by a dispersive full-parameter Jones matrix, as shown below:

$$J(\lambda) = R(-\theta)\begin{bmatrix} t_X(\lambda) & 0 \\ 0 & t_Y(\lambda) \end{bmatrix} R(\theta) = \begin{bmatrix} t_{XX}(\lambda) & t_{XY}(\lambda) \\ t_{YX}(\lambda) & t_{YY}(\lambda) \end{bmatrix} \quad (2)$$

where $R(\theta)$ is the rotation matrix of the meta-optics. $t_X$ and $t_Y$ are the complex response under two orthogonal linear polarizations, respectively, which are related the geometric parameters ($L$, $W$). The four wavelength-dependence complex components of Jones matrix represent the ability for dispersive full-parameter control. A more detailed derivation is provided in Supplementary Information S1. However, since the feature size is subwavelength, obtaining $t_X$ and $t_Y$ requires numerical simulation, which is both time consuming and nondifferentiable. To overcome this, we utilized a neural network (NN) to establish the connection between geometric parameters ($L$, $W$) of the meta-optics and their amplitude and phase responses across multiple wavelengths and polarizations.

The NN is trained on datasets comprising the phase ($\varphi$) and amplitude ($A$) of meta-atom, obtained through numerical simulations using the Finite Difference Time Domain (FDTD) method. Fig. 1c and Fig. 1d display the simulated results $\varphi_X$ and $A_X$ of meta-atom with sparse geometric parameters ($L$, $W$) under the X- polarization at discrete wavelengths from 400 nm to 700 nm. Here, W is fixed at 200 nm to simplify visualization. The training details and total simulated datasets are provided in Methods and Supplementary Information S1. Once trained, the NN's weights are stabilized, enabling rapid generation of the phase and amplitude of meta-optics, which is analytically expressed as $A_X, A_Y, \varphi_X, \varphi_Y = f_{NN}(L, W, \lambda)$. The regenerated phase $\varphi_X$ and amplitude $A_X$ of meta-atom are illustrated in Fig. 1e and 1f, with denser sampling and very close to the data obtained by FDTD, which proves the NN is well-trained. Thus, the NN-based dispersive full-parameter Jones matrix has achieved differentiability.

Considering an incident field $E_{in}(p, \lambda) = \begin{bmatrix} a_x(\lambda) & a_y(\lambda)e^{i\varphi(\lambda)} \end{bmatrix}^{-1}$, the modulation field $E_{ms}$ of different polarization states ($p$) and wavelengths ($\lambda$) is diffracted at specific regions ($x$, $y$, $z$) after the $E_{in}$ interacting with the NN-based dispersive full-parameter

Jones matrix $J$ of meta-optics. The forward simulator of the meta-optics has been constructed with differentiable properties and can be summarized in the following form:

$$E_{ms}(p,\lambda;x,y,z) = f_{propapa.} \langle J\{f_{NN}[L(x,y),W(x,y),\lambda],\theta(x,y)\},z \rangle \quad (3)$$

This forward simulator facilitates the extraction of gradient information with respect to the loss function for the geometric parameters $S$ of the meta-optics via automatic differentiation, represented as $\partial Loss/\partial S$. More details about gradient calculations are available in Supplementary Information S2. Subsequently, a new generation of $S$ is updated using the Adam optimizer. Compared with the traditional adjoint method, this approach eliminates the need for a full-wave simulation to obtain the perturbation of each parameter, thereby saving a substantial amount of computational resources. Gradient backpropagation though automatic differentiation is implemented in the open-source machine learning framework TensorFlow2. By continuously iterating the forward simulator and backpropagation process, optimization is conducted in the direction of gradient descent until the design criteria are met, ultimately outputting the final geometric parameters $S^{final}$ of meta-optics as shown in Fig. 1b.

**Dual-polarization color holography**

To demonstrate the superiority of end-to-end inverse design, we implemented dual-polarization color holography and compared its performance with that of sequential forward design. Fig. 2a illustrates the schematic of dual-polarization color holography, displaying two independent color holographic images on the same plane by switching two orthogonal linear polarization channels. These images are designed as "HNU" three-color letters and three primary color pictures, respectively. This functionality can be achieved by a birefringence meta-optics without rotation angle ($\theta = 0$). In this context, the dispersive Jones matrix of meta-optics can be simplified to $J(\lambda) = \begin{bmatrix} t_X(\lambda) & 0 \\ 0 & t_Y(\lambda) \end{bmatrix}$, where $t_x$ and $t_y$ associated with length ($L$) and width ($W$) of meta-optics, represent the dispersive complex amplitudes of meta-atom in two orthogonal linear polarizations. By adjusting the geometric parameters (*L*, *W*), we can control $t_x$ and $t_y$ to realize two color holographic images.

In sequential forward design, the only-phase holograms retrieved by computer-generated hologram (CGH) algorithms[48] are indicated by blue dots in Fig. 2b. Details of the sequential forward design are described in Supplementary Information S3. Due to the limited dispersive capabilities of a single meta-atom, complete phase coverage from 0 to $2\pi$ cannot be achieved simultaneously under RGB wavelengths, as illustrated by the yellow circle in Fig. 2b. This limitation results in significant phase errors across many regions at RGB wavelength channels during the process of minimizing error matching, significantly degrading the quality of the reconstructed holographic images. Visualizations of phase matching errors for each channel are presented in Supplementary Information Fig. S4.

By utilizing end-to-end inverse design, we can optimize complex amplitude hologram to better accommodate the limited dispersive capabilities, thereby enhancing the quality of reconstructed holographic images. The optimization goal is to minimize the loss function defined as the mean square error (MSE) and negative Pearson correlation (NPCC) between the intensities of actual holographic images and target images. Definitions of MSE and NPCC are provided in the Supplementary Information S2. Throughout the iterative loop to minimize loss function, the actual holographic images are continuously refined to closely approximate the target images. Consequently, no intermediate matching error is introduced, facilitating the achievement of superior image quality.

The holographic images of the meta-optics can be reconstructed using scalar simulation, vector simulation, and experimental measurement. Initially, in scalar simulated analysis, comparing two holograms of meta-optics with different sizes designed by two methods are show in Fig. 2c. The root means square error (RMSE) of each channel hologram achieved through end-to-end inverse design is significantly lower than that achieved through sequential forward design. Error bars in Fig. 2c represent the standard deviation across different sizes of meta-optics. Details of the comparison between two methods are shown in Supplementary Information S3.

We performed full-wave vector simulation of meta-optics to ensure the reliability of the design. Due to limited computing memory, the size of the meta-optics and the diffraction distance are scaled to 50 μm×50 μm and 75 μm, respectively. We also fabricated the samples of meta-optics designed by the two methods for experimental comparison. The size of meta-optics is set to 100 μm×100 μm and the diffraction distance is 150 μm. The top-view and oblique-view scanning electron microscope (SEM) images of the fabricated samples are shown in Fig. 2d. A color holographic optical path is constructed to test these samples. Details of the fabrication process and experimental optical path are described in Supplementary Information S6 and S7. Fig. 2e presents two target color images under two polarization channels. Fig. 2f-g compare the results of scalar simulation, vector simulation and experimental measurement for both methods. The similarity across three results confirms the accuracy and reliability of our proposed framework. The inferior results from vector simulation are attributed to the reduced size. For a quantitative assessment of image quality, we calculated the Peak Signal-to-Noise Ratio (PSNR) and Structural Similarity Index (SSIM) for each holographic image. The specific values of the metric are displayed in the upper left corner of the holographic images in Fig. 2f-g. These results show that end-to-end inverse design consistently outperforms the sequential forward design across all examined metrics. Detailed formulas of PSNR and SSIM are available in Supplementary Information S3.

**Tri-polarization color 3D holography**

To further utilize the design freedom of meta-atoms by considering rotation angles ($\theta \neq 0$), birefringent meta-optics can enable the polarization conversion channel. As mentioned in Eq. (2), the dispersive full-parameter Jones matrix encompasses three wavelength-dependent components: $t_{xx}(\lambda)$, $t_{yy}(\lambda)$, $t_{xy}(\lambda)=t_{yx}(\lambda)$. These three components can be extracted by alternately switching the linear polarizations of the input and output respectively. To enhance control over spatial dimensions, we established a color holographic reconstruction utilizing different polarization channels across various depth planes, facilitating tri-polarization color three-dimensional (3D) holography as depicted in Fig. 3a. When the incident and emergent light are X-polarized and Y-polarized, respectively, meta-optics displays the first color holographic image "rainbow" in depth plane $Z_1$. When both the incident and emergent light are Y-polarized, a second color holographic image "flower" is produced in depth plane $Z_2$. Similarly, when both are X-polarized, the third color holographic image "balloon" appears in depth plane $Z_3$. Given that the polarization conversion properties of birefringent meta-atom at different wavelengths are fixed, making the sequential forward design impractical. Instead, our proposed end-to-end inverse design effectively identifies suitable geometric parameters of meta-optics to manipulate phase and amplitude across multiple polarizations and wavelengths. We optimized a 150 μm×150 μm meta-optics to perform this function. Three depth planes were set at 150 μm, 175 μm, and 200 μm. Details of the implementation of tri-polarization color 3D holography are provided in Supplementary Information S4.

To improve the shape robustness of meta-optics to ensure high performance despite shape errors in fabrication, we incorporated a noise distribution into the geometric parameters of meta-optics during end-to-end inverse design. Consequently, the PSNR of three holographic images with noise tends to remain stable despite changes in shape error, as shown by the solid line in Fig. 3b. In contrast, the PSNR of the three color holographic images without noise is highly sensitive to variations in shape error. Details of shape robustness improvement are provided in the Supplementary Information S4. Subsequently, we fabricated the sample of meta-optics designed for tri-polarization color 3D holography. Fig. 3c displays top-view and oblique-view scanning electron microscope (SEM) images of the fabricated sample. By testing this sample, we captured three color holographic images displayed in the Fig. 3e. Compared with the target image in Fig. 3d, the captured holographic images exhibit considerable quality, with their PSNR and SSIM displayed in the upper left corner of Fig. 3e. Notably, there is no crosstalk between each color image, confirming the independence of each polarization channel. Moreover, the nine holographic images, corresponding to three wavelengths and three polarizations across three depth planes as shown in Fig. 3f, are also very clear and sharp. However, the proximity of the blue light wavelength (450 nm) to the unit period (400 nm) of the meta-optics makes the blue channel more sensitive to fabrication and experimental errors, resulting in slight crosstalk. This issue can be mitigated by adjusting the configuration of the meta-optics.

**Arbitrary polarized spectral multi-information processing**

Simultaneously controlling arbitrary polarization and wavelength could significantly extend the potential applications in the realms of polarized spectral multi-information processing. However, the number of independent polarization channels remains limited due to the inherent constraints in the degrees of freedom of the Jones matrix[35]. Although superimposing polarization components enables the synthesis of arbitrary polarization states, designing meta-optics with a single rectangular meta-atom to manipulate these states across multiple wavelengths is still impractical using sequential forward design. To demonstrate the universality of our end-to-end inverse design, we employed it to design meta-optics with a single rectangular meta-atom for controlling arbitrary polarizations and wavelengths. This design facilitates intelligence application in polarized spectral multi-information processing. Fig. 4a illustrates the schematic of this processing approach. The meta-optics directs incoming light of any given polarization and wavelength into specific spatial regions corresponding to certain polarizations and wavelengths, thereby enabling the acquisition of multi-dimensional information.

For the multi-information display, we implemented a meta-optics device capable of displaying six holographic images at different locations, each corresponding to six arbitrary polarizations across three wavelengths. The polarizations include X-linear polarization (XLP), Y-linear polarization (YLP), left-handed elliptical polarization (LEP), right-handed elliptical polarization (REP), left circular polarization (LCP), and right circular polarization (RCP). The wavelengths employed are $\lambda_1$=635 nm, $\lambda_2$=532 nm, and $\lambda_3$=450 nm. Each combination of wavelength and polarization uniquely corresponds to a distinct holographic image: an Arabic numeral "1" for $\lambda_1$ with XLP, "2" for $\lambda_2$ with LEP, "3" for $\lambda_3$ with LCP, "4" for $\lambda_1$ with YLP, "5" for $\lambda_2$ with REP, and "6" for $\lambda_3$ with RCP. Fig. 4b presents the experimental captured reconstructed holographic images for six polarizations at each of the three wavelengths, with the specific polarization state indicated in the upper left corner of each image. The experimental results confirm the successful reconstruction of predetermined holographic images at the targeted polarizations and wavelengths, demonstrating good image quality. We calculated the correlation coefficients for each image and plotted the correlation coefficient matrix for the six channels, as shown in Fig. 4c. The independence of the holographic images across each channel verifies that the designed meta-optics can effectively distinguish between polarization and spectral information based on the specific values of the images displayed in different spatial positions.

To highlight broader applications in multi-information processing, we utilized a multi-focal metalens for polarized spectral detection by modifying the objective function in our end-to-end inverse design (details provided in Supplementary Information S2). Fig. 4d illustrates the focus characterization of the six-focal metalens under six polarizations and three wavelengths measured in optical testing. Light with different polarization states and wavelengths is divided into distinct regions on the imaging plane. We also measured the focal intensity of six polarizations within the visible spectrum (400 nm -700 nm) using ten discretely measured color filters at various wavelengths. Fig. 4e plots the sorting efficiency curve as the spectrum varies, highlighting that the light intensity is strongest at the target focal points for the designed wavelengths. This demonstrates the capability for spectral detection in six polarization channels. For the polarization detection, we measured the focus intensity with different state of polarizations (SoPs) at three wavelengths. Fig. 4f illustrates the polarization contrast curve with the variation of SoPs. As the SoPs change, the polarization contrast ranges from 1 to -1, indicating the polarization detection capability at three wavelength channels. The sorting efficiency in the spectral dimension and polarization contrast in polarization dimension demonstrate that our designed meta-optics can simultaneously detect the multi-information of polarization and spectrum. The focal images of different wavelengths and SoPs captured in the experiment tests, along with the definitions of sorting efficiency and polarization contrast, are detailed in Supplementary Information S5.

To show the incoherent imaging performance of our meta-optics, we measured the imaging of the resolution target illuminated by mercury lamps in the optical test. The third level of the resolution chart was imaged at different positions, each with varying polarization and wavelength, as illustrated in Fig. 4g. The absence of crosstalk in the non-target region highlights the effectiveness of the designed meta optics in detecting polarization and spectral information. Below each image, we plot the horizontal and vertical intensity profiles of each imaging pattern, revealing high-resolution details that indicate superior imaging quality. The details of the optical setup used for testing the focus characterization and imaging performance are provided in Supplementary Information S7. The results presented above indicate that our framework applies to both coherent and incoherent light, significantly facilitating the field of multi-dimensional computational displaying and imaging.

The cases demonstrated above focus primarily on RGB wavelengths to simplify experimental measurements, yet this is not the upper limit of our proposed framework. The neural network-based dispersive full-parameter Jones matrix depicted in Fig. 1b can generate phase and amplitude across a broader spectrum of wavelengths. We further expanded the demonstration for more wavelengths, polarizations, and planes multiplexing, as detailed in Supplementary Information S8 and S9. Additionally, this differentiable framework is compatible with the training paradigms of deep learning models commonly used in computer vision and image processing, thus promoting the integration of meta-optics with computational imaging[49] and optoelectronic diffractive network[50]. Meta-optics for spectral and polarization information process can surpass the low-dimensional limitations of traditional photonic sensors. This advanced capability facilitates the encoding of high-dimensional information to a greater extent and allows for decoding through backend processing algorithms, such as multidimensional fusion imaging that integrates spectral polarization, phase, and depth[51,52].

## Discussion

We developed a versatile, general end-to-end inverse design framework for non-interleaved meta-optics to enable full-parameter control across multiple wavelengths at each spatial location. Compared to sequential forward design, our proposed framework gained a substantial improvement of the quality in dual-polarization color holography. In situations where sequential forward design is impractical, we achieved tri-polarization color 3D holography experimentally for the first time, including nine independent channels to display three color images across three linear polarizations and three depth planes. Moving beyond linear polarization, we combined six arbitrary polarizations with three wavelengths to create two types of meta-optics: six-channel meta-holograms and six-focal metalens. The polarization and spectral sensitivity of these meta-optics were experimentally verified, showcasing their capacities for multi-information processing. This differentiable and system-level end-to-end inverse design elevates meta-optics to a new standard in comprehensive full-parameter light field manipulation, facilitating a synergy between AI and meta-optics. Ultimately, our efforts aim to accelerate the development of meta-optics for integrated multi-information display, imaging, communication, and to extend their applications in multi-modal sensing.

## Methods

**Numerical simulation**

The meta-atom and meta-optics were simulated using Lumerical FDTD. The period of meta-atom was set as 400 nm, periodic boundary conditions were adopted in the x and y directions, and the perfect matching layer was adopted in the y direction. Two-dimensional parameter scanning was performed on a single element with a length and width of 60 nm ~ 340 nm and a fixed height of 1000 nm in visible light (400 nm~700 nm) and X- polarization. The refractive index of $SiO_2$ is 1.46, and the refractive index of $TiO_2$ is measured by ellipsometer. The phase and amplitude transmittance of meta-atom with different geometric parameters are shown in Fig. S1. For full-wave simulation of meta-optics, we only simulated 125×125 pixels (50 μm×50μm) because the limited computing memory. The boundary condition was set to perfect matching layers (PML) in x, y, and z directions.

**Neural network training**

The neural network in forward simulator comprises an input layer, four hidden layers (64-256-256-64 neurons) and an output layer (Supplementary Information Fig. S2a). ReLu is used as the activation function. The input layer accepts the meta-atom geometric parameters ($L, W$) and the incident light wavelength ($\lambda$). Incorporating wavelength as an input avoids low-to-high dimensional mapping, facilitating network convergence. The output layers are $sin(\varphi_X), cos(\varphi_X), sin(\varphi_Y), cos(\varphi_Y), A_X, A_Y$, with sinusoidal projection for phase conversion to prevent phase wrapping-related convergence issues. We used the data simulated in Lumerical FDTD for network training and testing, with a ratio of 8:2. The network was trained using stochastic gradient descent to minimize the mean absolute error (MAE) using Adam optimizer with a learning rate of 0.0001, and 30,000 epochs in TensorFlow2. The loss curve (Supplementary Information Fig. S2b) demonstrates synchronized convergence of training and validation losses with minimal differences.

**Sample fabrication**

The meta-optics samples were fabricated by electron beam lithography (EBL) along with etching techniques. Initially, a 1000 nm electron beam inhibitor layer of polymethyl methacrylate (PMMA) was spin-coated on a transparent silica substrate coated with an ITO film at a rotational speed of 2000 rpm and then baked on a hot plate at 180°C for 4 minutes. The pattern was subsequently exposed using electron beam lithography (EBL) at 100 kV voltage and 200 pA current. Following this, the sample was immersed in a mixture of methyl isobutyl ketone (MIBK) and isopropyl alcohol (IPA) (MIBK: IPA = 1:3) for 1 minute at a low temperature, and then fixed in IPA for 1 minute. Afterward, a 230 nm thick $TiO_2$ was deposited on the exposed area using atomic layer deposition (ALD), and the excess $TiO_2$ on the top of the sample was removed by ion beam etching (IBE). Finally, the reactive ion etching system (RIE) was used to remove the photoresist by injecting $O_2$ gas, leaving a high-aspect-ratio $TiO_2$ nanostructure on the substrate.

**Experiment setup**

To verify the performance of meta-optics for dispersive full-parameter manipulation, we designed three optical setups to characterize color holography, focusing and imaging, respectively. Detailed characterizations of these optical experimental setup are shown in the Supplementary Information S7.

**Acknowledgments**

We acknowledge the financial support by the National Natural Science Foundation of China (Grant Nos. 62275078, 62105120), Natural Science Foundation of Hunan Province of China (Grant No. 2022JJ20020), Science and Technology Innovation Program of Hunan Province (Grant No. 2023RC3101), and Shenzhen Science and Technology Program (Grant No. JCYJ20220530160405013).

**Author contributions**

H.C. and Y.H. proposed the idea. H.C. carried out the design, simulation, optimizations, led the experiment, and authored manuscript. H.C., Q.W. and Q.X. fabricated all the samples. H.C., X.O., S.L., and Y.J. conceived and performed the measurements. H.C., H.Y., C.W.Q., and H.D. discussed the results and revised the manuscript. Y.H. and H.D. supervised the overall project. All authors confirmed the final paper.

**Competing interests**

The authors declare no competing interests.

Freedom Metasurfaces. *Adv. Funct. Mater.* **30**, 1910610 (2020).

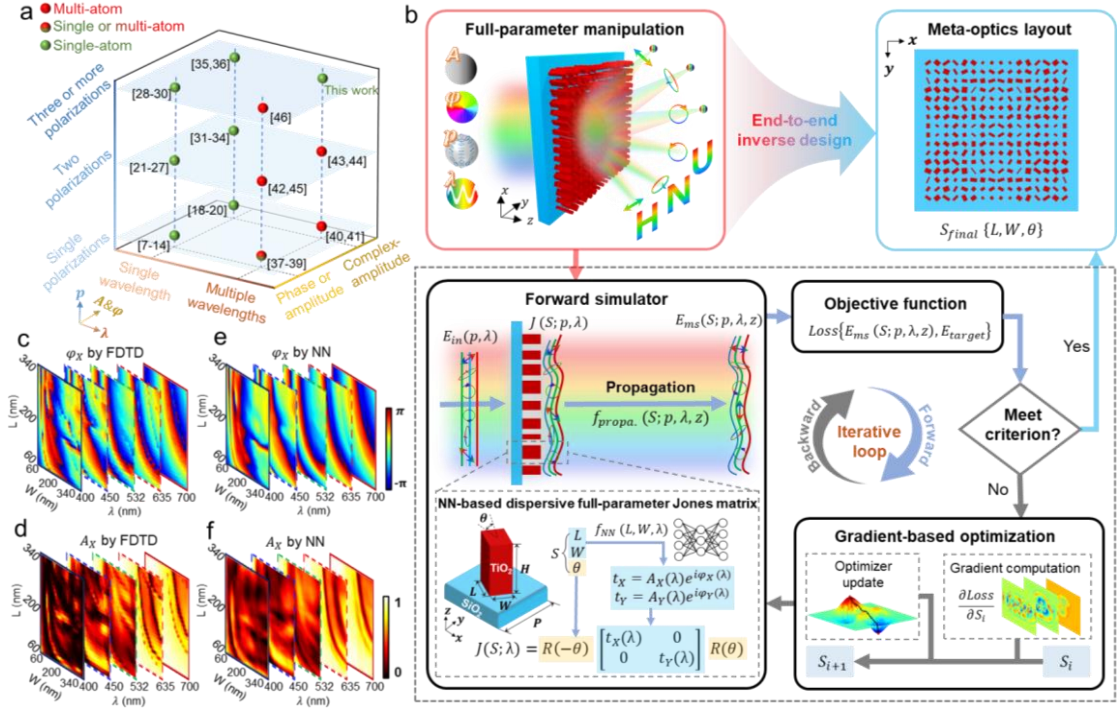

**Fig. 1 | Dispersive full-parameter control of meta-optics. a,** A summary of multi-parameter light filed control with single-layer meta-optics in the literatures. **b,** End-to-end inverse design enables the direct output of layouts for fabricating meta-optics with full-parameter control. It includes a forward simulator featuring a neural network-based dispersive full-parameter Jones matrix and Fourier propagation, as well as backward optimization based on an objective function, gradient backpropagation and an iterative loop. **c, d,** Simulated sparse data for the phase $\varphi_X$ and amplitude $A_X$ of the rectangular meta-atom with no rotation (θ=0), under X- polarization at wavelengths of 400-700 nm by FDTD method. **e, f,** Dense data regenerated by the pre-trained NN, compared with the sparse data by FDTD.

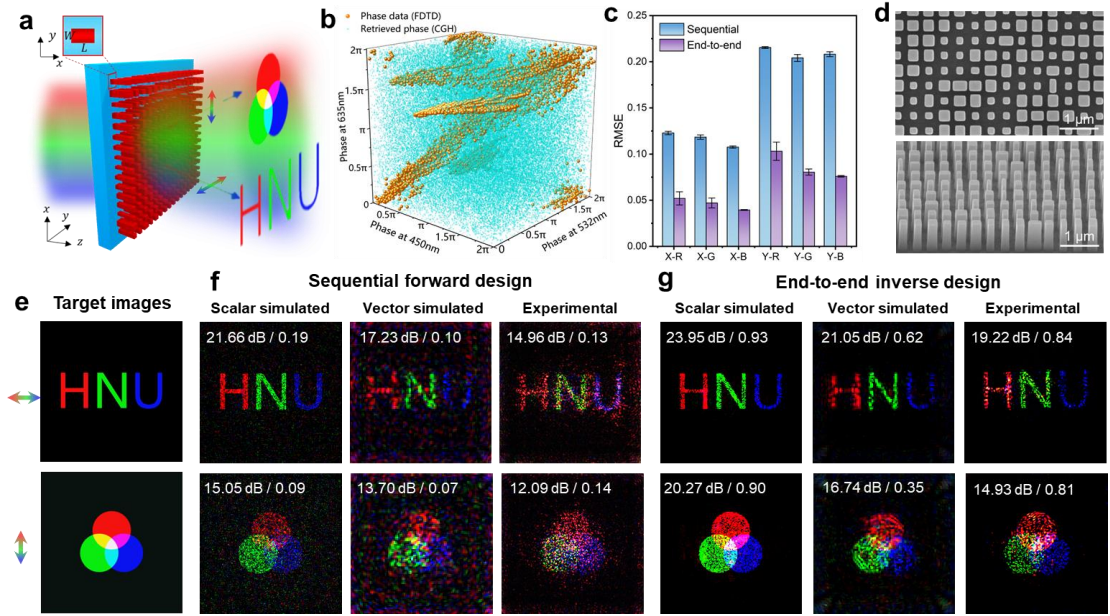

**Fig. 2 | Comparison of the dual-polarization color holography by sequential forward design and end-to-end inverse design. a,** Schematic of dual-polarization color holography. **b,** Distribution of phase data simulated by FDTD and the retrieved phase obtained by CGH at RGB wavelengths. **c,** Comparison of the RMSE for simulated holographic images with averages of different sizes, between the sequential forward (bule bars) and end-to-end inverse (purple bars) design. Error bars represent the standard deviation across different sizes. **d,** Top-view (up) and oblique-view (bottom) of the scanning electron microscope (SEM) images of the fabricated sample. Scale bar: 1 µm. **e,** Target color images for X- and Y- polarizations. **f, g,** Calculated, simulated, and experimentally measured holographic images achieved by sequential forward design and end-to-end inverse design.

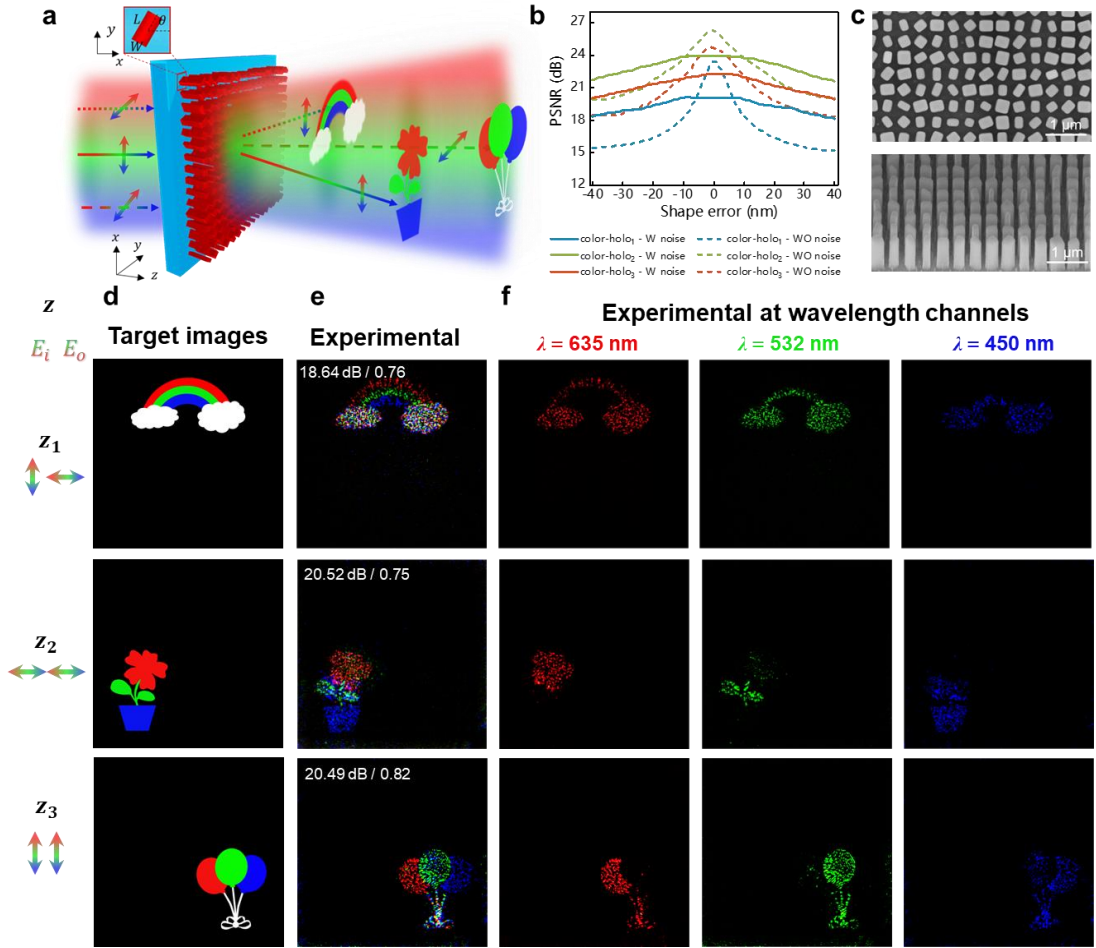

**Fig. 3 | Demonstration of tri-polarization color 3D holography. a,** Schematic of tri-polarization color 3D holography, illustrating three colors holographic images at different depth planes that can be switched with two orthogonal polarization inputs. **b,** PSNR values of scalar simulations for three colors holographic images as a function of shape error with (W) and without (WO) noise during optimization. **c,** Top-view (up) and oblique-view (bottom) of the scanning electron microscope (SEM) images of the fabricated sample. Scale bar: 1 μm. **d, e,** Target and experimentally measured color holographic images of "rainbow", "flower", and "ballon" under different input/output polarizations, with each channel arranged at a distinct depth plane. **f,** Nine individual holographic images experimentally measured at each of three wavelengths and three polarizations in three depths planes.

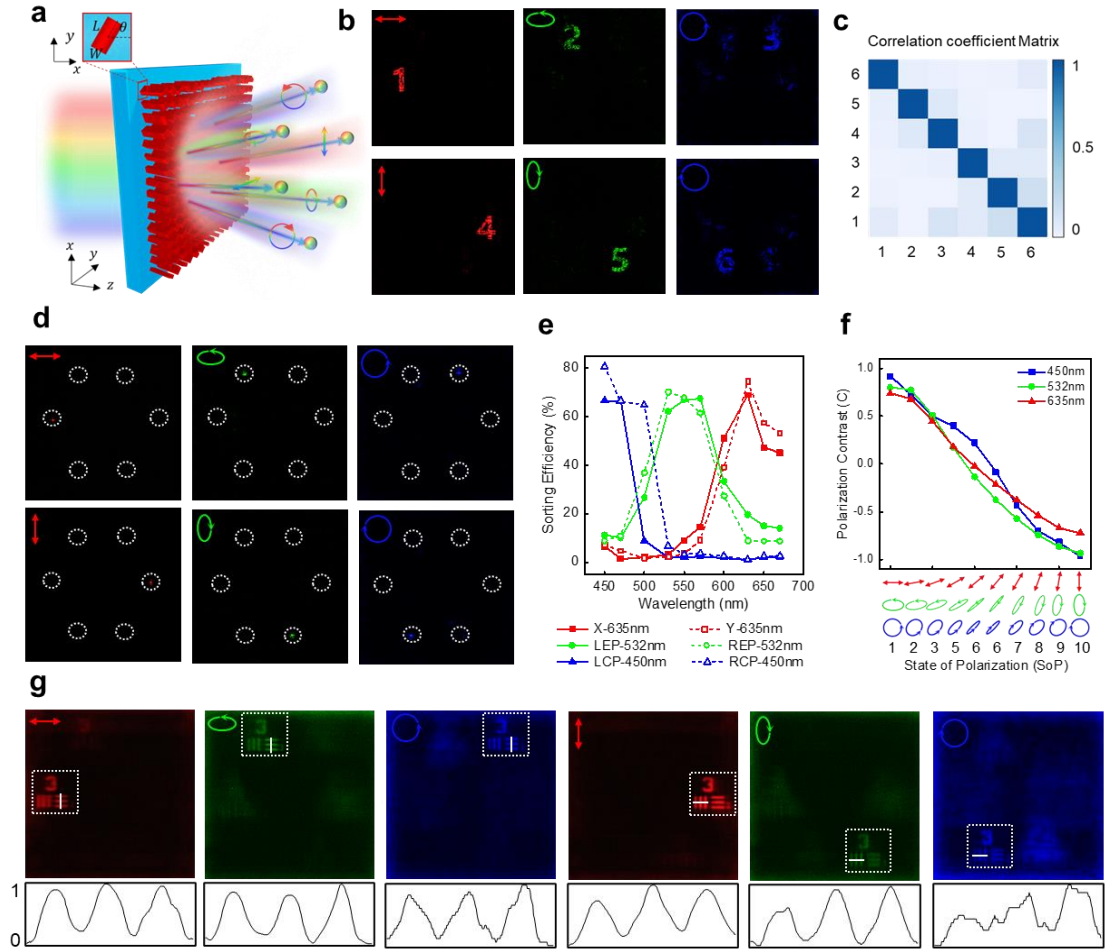

**Fig. 4 | Demonstration of arbitrary polarized spectral multi-information processing. a,** Schematic of arbitrary polarized spectral multi-information processing. **b,** Experimentally measured holographic images displaying Arabic numerals "1" to "6" for six channels meta-holograms. **c,** The correlation coefficient matrix for six holographic images. **d,** Experimentally measured focusing characteristics at different wavelengths and polarizations for six-focal metalens. **e,** Experimentally sorting efficiency as a function of wavelength. **f,** Experimentally polarization contrast as a function of States of Polarization (SoPs). **g,** Experimentally measured imaging of a resolution target at various wavelengths and polarizations. Below each image, the cross-sectional curve of the white tangent line is shown.